\documentclass{PoS}
\usepackage{amssymb,amsmath,graphicx,bm}
\newcommand{\xB}{x_{\scriptscriptstyle B}}
\newcommand{\sT}{{\scriptscriptstyle T}}
\renewcommand{\d}{\mathrm{d}}

\title{Probing Gluon TMDs at a Future EIC}

\ShortTitle{Probing Gluon TMDs at a future EIC}

\author{\speaker{Cristian Pisano}\\
        Dipartimento di Fisica, Universit\`a di Pavia and INFN, Sezione di Pavia \\
        Via Bassi 6, I-27100 Pavia, Italy \\
        E-mail: \email{cristian.pisano@unipv.it}}

\author{Dani\"el Boer\\
        Van Swinderen Institute for Particle Physics and Gravity, University of Groningen\\
        Nijenborgh 4, 9747 AG Groningen, The Netherlands\\
        E-mail: \email{d.boer@rug.nl}}

\author{Piet J.~Mulders\\
       Nikhef and Department of Physics and Astronomy, VU University Amsterdam\\
        De Boelelaan 1081, NL-1081 HV Amsterdam, The Netherlands\\
        E-mail: \email{mulders@few.vu.nl}}
        
\author{Jian Zhou\\
        School of physics and Key Laboratory of Particle Physics and Particle Irradiation (MOE)\\
        Shandong University, Jinan, Shandong 250100,China\\
        E-mail: \email{jzhou@sdu.edu.cn}}

\abstract{
Gluon TMDs can be accessed through the analysis of azimuthal asymmetries for heavy quark pair and dijet production in electron-proton collisions, similarly to the way quark TMDs are commonly extracted from semi-inclusive deep-inelastic scattering data. We calculate the upper bounds for these observables, showing in which kinematic regions they are large enough to be measured in future experiments at an Electron-Ion Collider. Moreover,  we study  their behavior in the small-$x$ region, adopting a McLerran-Venugopalan model for unpolarized and linearly polarized gluon distributions. By comparison with related observables at RHIC and LHC, we expect to gather information on the process dependence of the gluon TMDs and to test our prediction of a sign change of the gluon Sivers function and two other $T$-odd gluon distributions. }

\FullConference{QCD Evolution 2016\\
		May 30-June 03, 2016\\
		National Institute for Subatomic Physics (Nikhef), Amsterdam}

\begin{document}

\section{Introduction}
Transverse momentum dependent parton distribution functions (TMDs) encode fundamental information on the intrinsic motion of partons and the correlation between their spins and momenta, providing a full three-dimensional picture of the proton in momentum space. While much effort has been made in order to extract  quark TMDs, mainly from low energy data on semi-inclusive deep inelastic scattering (SIDIS), almost nothing is known experimentally about gluon TMDs.

 In case of unpolarized protons, in addition to the unpolarized gluon TMD, the distribution of linearly polarized gluons is of special interest~\cite{Mulders:2000sh}: it corresponds to an interference between $+1$ and $-1$ gluon helicity states and, being $T$-even, it can in principle be nonzero even in absence of initial (ISI) and final state interactions (FSI).  However this distribution, like all other TMDs, is affected by such interactions that  can render it process dependent.
For a transversely polarized proton, exactly in the same way as for quarks, it is possible to define a  $T$-odd Sivers function for unpolarized gluons~\cite{Sivers:1989cc,Boer:2015ika}, which describes their azimuthal distribution and might give rise to sizable left-right asymmetries.  Furthermore, there are two other independent $T$-odd gluon TMDs: both of them are helicity-flip distributions of linearly polarized gluons. 

In this contribution to the proceedings, after providing the definition of gluon TMDs in terms of QCD operators, we 
show how they can be probed directly by looking at transverse single spin and azimuthal asymmetries  in both heavy quark pair and dijet production in electron-proton collisions~\cite{Boer:2016fqd,Boer:2010zf,Pisano:2013cya}. Moreover, we 
discuss their process dependence and behavior in the small-$x$ regime.  A test of our predictions would be possible at a future Electron-Ion Collider (EIC)~\cite{Boer:2011fh} where both the small and large $x$ regions can be accessed. We suggest related measurements in proton-proton collisions that could be performed at RHIC and the LHC, where a possible
fixed target experiment, called AFTER@LHC~\cite{Brodsky:2012vg}, would allow to investigate the structure of  transversely polarized protons as well. A consistent picture among the various processes, at large  and small $x$, will confirm our understanding of transverse spin effects and the TMD formalism in general.
 
\section{Definition of gluon TMDs}
We consider a gluon that carries momentum $p$ inside a proton with momentum $P$ and spin vector $S$. If we introduce a light-like vector $n$ conjugate to $P$, the longitudinal and transverse components of gluon momentum can be defined via the Sudakov decomposition $p = x\,P + p_\sT + p^- n$. Similarly, the spin vector of the proton can be written as 
\begin{equation}
S^\mu = \frac{S_L}{M_p}\, \bigg (  P^\mu -  \frac{M_p^2}{P\cdot n}\, n^\mu\bigg )  + S_\sT^\mu\,,
\end{equation}
where $S_L^2 + \bm S_\sT^2= 1$ and $M_p$ is the proton mass. Gluon TMDs are defined through the following correlator~\cite{Mulders:2000sh}, 
\begin{align}
\label{GluonCorr}
 {\Gamma}_g^{\mu\nu}(x,\bm p_\sT )
& =  \frac{n_\rho\,n_\sigma}{(P{\cdot}n)^2}
{\int}\frac{\d(\xi{\cdot}P)\,\d^2\xi_\sT}{(2\pi)^3}\ e^{ip\cdot\xi}\, \langle P,
S|\,{\rm{Tr}} \big[\,F^{\mu\rho}(0)\, U_{[0,\xi]} F^{\nu\sigma}(\xi)\,U^\prime_{[\xi, {0}]}\,\big] \,|P, S \rangle\,\big\rfloor_{\text{LF}}\,,
\end{align}
with the gluon field strengths $F^{\mu \nu}(0)$ and $F^{\nu \sigma}(\xi)$ evaluated at a fixed light-front (LF) time $\xi^+ =\xi{\cdot}n=0$. The process dependent Wilson lines or gauge links $U_{[0,\xi]}$  and $U^\prime_{[0,\xi]}$ in Eq.~(\ref{GluonCorr}) encode the effects of ISI/FSI and are needed to ensure gauge invariance. Introducing  the symmetric and antisymmetric transverse projectors,  $g^{\mu\nu}_{\sT} = g^{\mu\nu} - P^{\mu}n^{\nu}/P{\cdot}n-n^{\mu}P^{\nu}/P{\cdot}n$ and 
$\epsilon_\sT^{\mu\nu}  = \epsilon_\sT^{\alpha\beta\mu\nu} P_\alpha n_\beta/P\cdot n$  (with $\epsilon_\sT^{1 2} = +1$),  the correlator 
for an unpolarized ($U$) can be parametrized  in terms of gluon TMDs as follows,
\begin{align}
 {\Gamma}_U^{\mu\nu}(x,\bm p_\sT )  \,= \, \frac{x}{2}\,\bigg \{-g_\sT^{\mu\nu}\,f_1^g (x,\bm p_\sT^2) +\bigg(\frac{p_\sT^\mu p_\sT^\nu}{M_p^2}\,
    {+}\,g_\sT^{\mu\nu}\frac{\bm p_\sT^2}{2M_p^2}\bigg) \,h_1^{\perp\,g} (x,\bm p_\sT^2) \bigg \}\,,
\label{eq:PhiparU}
\end{align}    
where $f_1^g$  is the unpolarized gluon TMD and $h_1^{\perp\,g}$ is the distribution of linearly polarized gluons.   Analogously, for a transversely polarized ($T$) proton,    
  \begin{align}
 {\Gamma}_T^{\mu\nu}(x,\bm p_\sT )  \,=\, & \frac{x}{2}\,\bigg \{g^{\mu\nu}_\sT\,
    \frac{ \epsilon^{\rho\sigma}_\sT p_{\sT \rho}\, S_{\sT\sigma}}{M_p}\, f_{1T}^{\perp\,g}(x, \bm p_\sT^2) + i \epsilon_\sT^{\mu\nu}\,
    \frac{p_\sT \cdot S_\sT}{M_p}\, g_{1T}^{g}(x, \bm p_\sT^2) \nonumber \\
    & \,  + \,  \frac{p_{\sT \rho}\,\epsilon_\sT^{\rho \{ \mu}p_\sT^{\nu \}}}{2M_p^2}\,\frac{p_\sT\cdot S_\sT }{M_p} \, h_{1 T}^{\perp\,g}(x, \bm p_\sT^2)\,- \,\frac{p_{\sT \rho} \epsilon_\sT^{\rho \{ \mu}S_\sT^{\nu \}}\, + \,
      S_{\sT\rho} \epsilon_\sT^{\rho \{ \mu } p_\sT^{\nu \}}}{4M_p} \, h_{1T}^{g}(x, \bm p_\sT^2) \,\,\bigg \}\,
\label{eq:PhiparT}
\end{align}
where $f_{1 T}^{\perp\,g}$ is the $T$-odd gluon Sivers function. The distributions $h_{1T}^{\perp\,g}$ and $h_{1T}^{g}$, which  are $T$-odd as well, are named 
in analogy with the quark ones~\cite{Meissner:2007rx}, although quark and gluon TMDs with the same name have different properties under symmetry operations. 
 The $h$ functions for quarks are chiral-odd and do not mix with the chiral-even $h$ functions for gluons. These distributions have to satisfy the following positivity bounds~\cite{Mulders:2000sh}
\begin{align}
\frac{\vert \bm p_\sT \vert }{M_p}\, \vert f_{1T}^{\perp \,g}(x,\bm p_\sT^2) \vert  \le   f_1^g(x,\bm p_\sT^2)\,,\quad
\frac{\vert \bm p_\sT \vert }{M_p}\, \vert h_{1}^g(x,\bm p_\sT^2) \vert  \le   f_1^g(x,\bm p_\sT^2)\,,\quad
\frac{\vert \bm p_\sT \vert^3}{2 M_p^3}\, \vert h_{1T}^{\perp \,g}(x,\bm p_\sT^2) \vert & \le   f_1^g(x,\bm p_\sT^2)\,,
\label{eq:bound}
\end{align}
where we have defined the combination $h_1^g \equiv h_{1T}^g + {\bm p_\sT^2}/({2 M_p^2})\,  h_{1T}^{\perp\,g}$,
which is not related to the well-known transversity distribution $h_1^q$ for
quarks that has no analogue in the gluon sector. This notation is justified by the fact that 
both $h_1^q$ and $h_1^g$ denote helicity
flip distributions. Moreover, we will show that they  can give rise to the same angular dependences in different types of processes.
However, $h_1^q$ is chiral-odd, $T$-even, and survives after transverse
momentum integration, whereas $h_1^g$  is chiral-even, $T$-odd, and vanishes when
integrated over transverse momentum. 

\section{Azimuthal asymmetries for heavy quark pair  and dijet production in DIS}
We first consider the process
$ e(\ell) + p(P,S) \to e(\ell^{\prime}) + {Q}(K_1) +  {\overline Q}(K_2)+ X$, 
where the proton has polarization vector $S$ and the other particles are unpolarized. We fix a reference frame such that 
the  $\hat z$-axis is along the direction of the proton and the virtual photon exchanged in the reaction, while the azimuthal angles are measured w.r.t.\ the lepton plane ($\phi_{\ell}=\phi_{\ell^\prime}=0$). In the so-called correlation limit, the two heavy quarks with transverse momenta $K_{i \perp}$ ($i=1,2$) such that $K_{i\perp}^2 = -\bm K_{i\perp}^2$,  are almost back to back in the plane orthogonal to $\hat z$.  Hence,  if we define $q_\sT \equiv K_{1\perp} + K_{2\perp}$ and $K_\perp \equiv (K_{1\perp} -K_{2\perp})/2$, we have $\vert q_\sT \vert \ll \vert K_\perp\vert $. Denoting by $\phi_S$, $\phi_\sT$ and $\phi_\perp$ the azimuthal angles of the three-vectors $\bm S_\sT$,  $\bm q_\sT$ and  $\bm K_\perp$, the cross section at leading order in perturbative QCD can be written in the following form
\begin{equation}
\frac{\d\sigma}
{\d y_1\,\d y_2\,\d y\,\d\xB\,\d^2\bm{q}_{\sT} \d^2\bm{K}_{\perp}} \equiv \d\sigma (\phi_S, \phi_\sT,\phi_\perp) =    \d\sigma^U(\phi_\sT,\phi_\perp)  +  \d\sigma^T (\phi_S, \phi_\sT,\phi_\perp)  \,,
\label{eq:cs}
\end{equation}
with $y_i$ being the rapidities of the heavy quarks, $y$ is the inelasticity variable, $x_B=Q^2/(2 P\cdot q)$, where $Q^2= -q^2= -(\ell-\ell^\prime)^2$ is the virtuality of the photon.  The full expressions for $\d\sigma^U$ and $\d\sigma^T$ are given in Ref.~\cite{Boer:2016fqd}.
The azimuthal modulations that provide access to 
$h_1^{\perp\,g}$ can be singled out by means of the following observables~\cite{Boer:2010zf,Pisano:2013cya}
 \begin{equation}
\vert \langle {\cal W}( \phi_\sT,\phi_\perp) \rangle \vert = \left| \frac{\int 
\d \phi_\perp\d \phi_\sT
\,{\cal W} (\phi_\sT,\phi_\perp) \, \d\sigma}{\int \d \phi_\perp \d \phi_\sT
\, \d\sigma}\right| = \frac{\bm q_\sT^2}{2 M^2}\,\frac{|h_1^{\perp\, g}(x,\bm p_\sT^2 )|}{ f_1^{g}(x,\bm p_\sT^2 )} \, {\cal F}^{\cal W}(y,z,Q^2,M_\perp^2)\,,
\label{eq:lpg}
\end{equation}
with ${\cal W}(\phi_\sT,\phi_\perp) =  \cos 2 \phi_\sT$, $ \cos 2 (\phi_\sT-\phi_\perp)$. Moreover, $z= P\cdot K_1/P\cdot q$ and $M_\perp^2 =  M_Q^2 + \bm K_{\perp}^2 $, where $M_Q$ is the mass of the heavy quark, while the functions $ {\cal F}^{\cal W}$ are presented explicitly in Ref.~\cite{Boer:2016fqd}.

\begin{figure}[t]
\begin{center}
 \includegraphics[angle=0,width=0.48\textwidth]{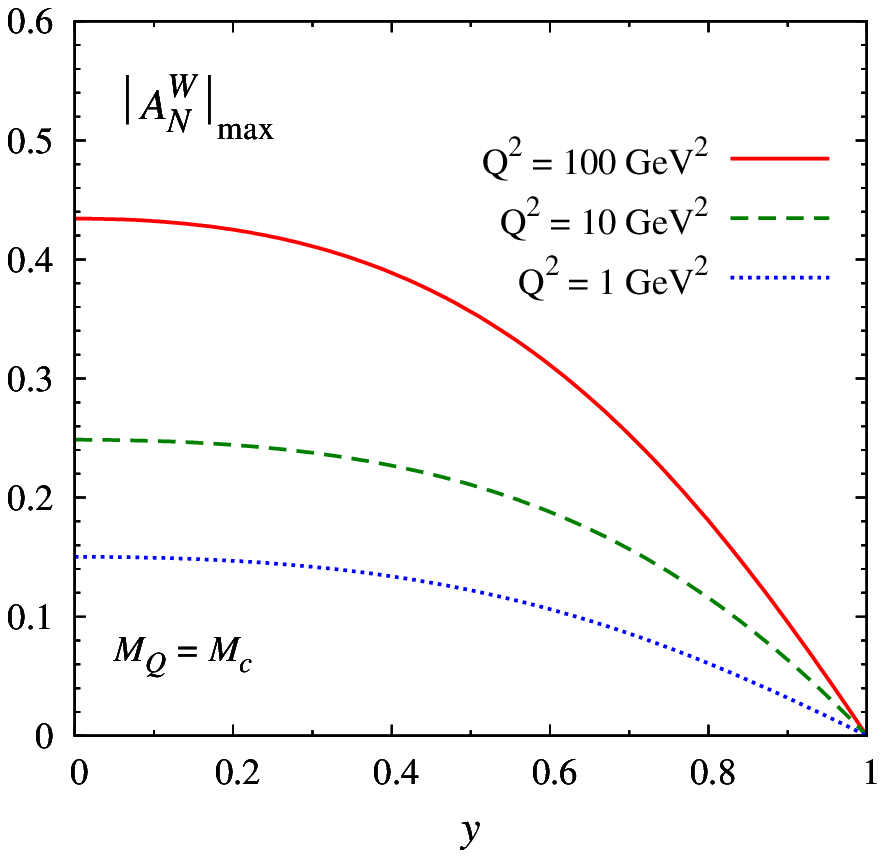}
 \includegraphics[angle=0,width=0.48\textwidth]{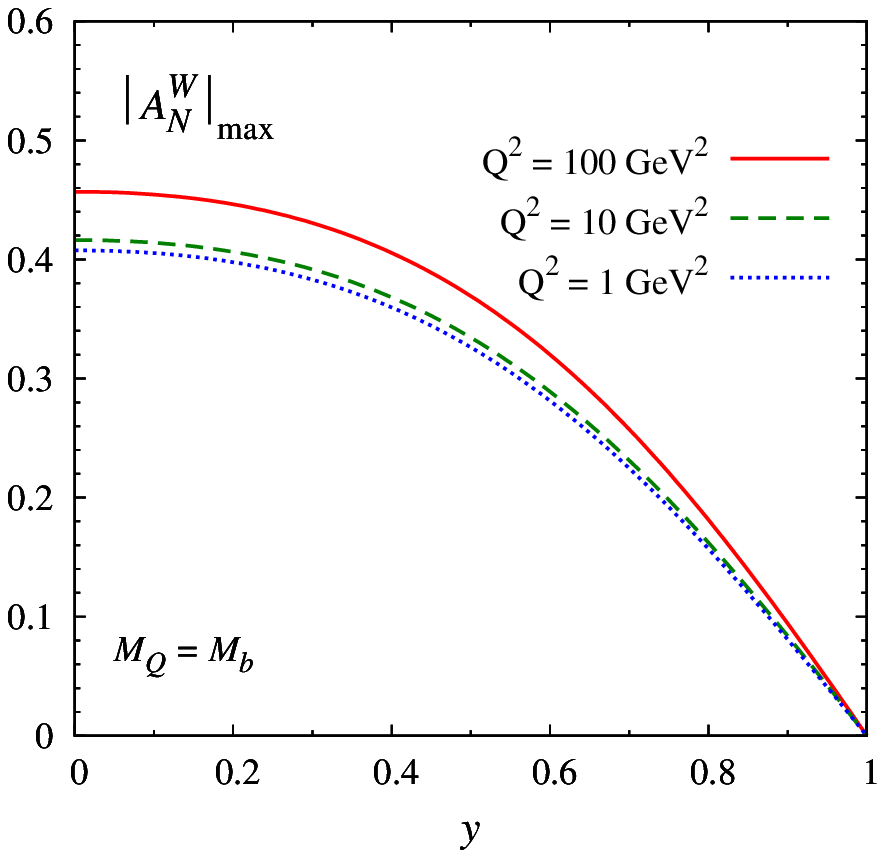}
 \caption{Upper bounds $\vert A_N^W\vert_{\rm max}$ on the asymmetries $A_N^{\sin(\phi_S+\phi_\sT)}$ and $A_N^{\sin(\phi_S-3\phi_\sT)}$  for the processes $e p\to e^\prime c \,\bar{c}\, X$ (left panel) and  $e p\to e^\prime b \,\bar{b}\, X$ (right panel). Estimates are given as a function of $y$, at  $z=0.5$ and fixed  values of  $Q^2$ and  $K_\perp (> 1$ GeV), the latter chosen in such a way that the asymmetries are maximal.}
\label{fig:ANW_y}
\end{center}
\end{figure}
\begin{figure}[t]
\begin{center}
 \includegraphics[angle=0,width=0.48\textwidth]{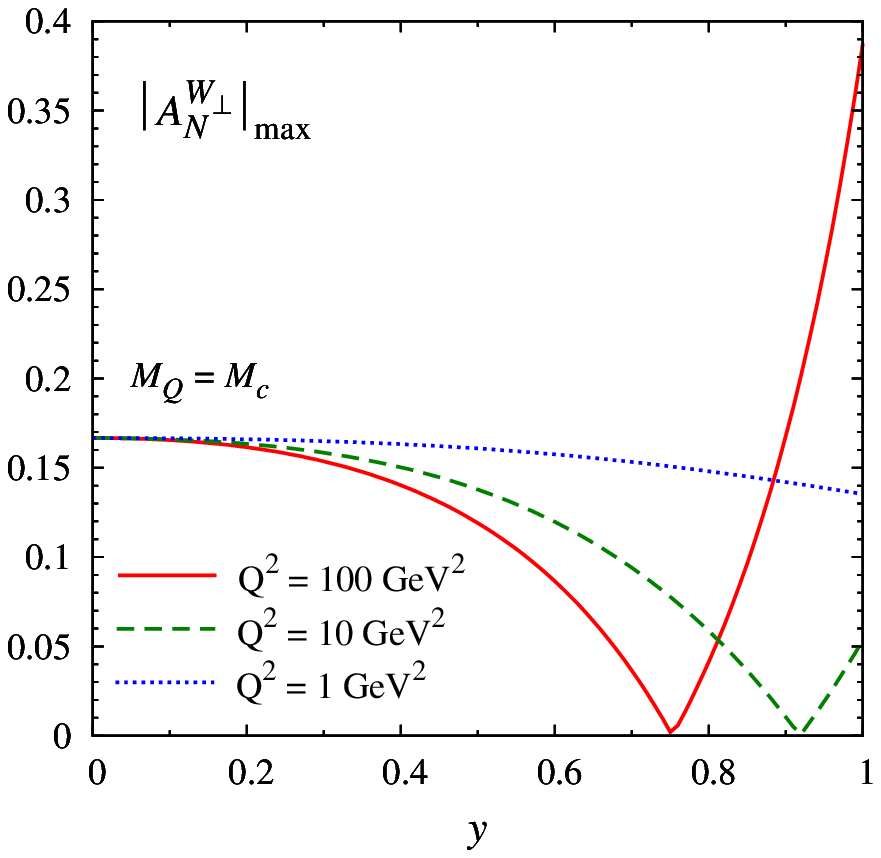}
 \includegraphics[angle=0,width=0.48\textwidth]{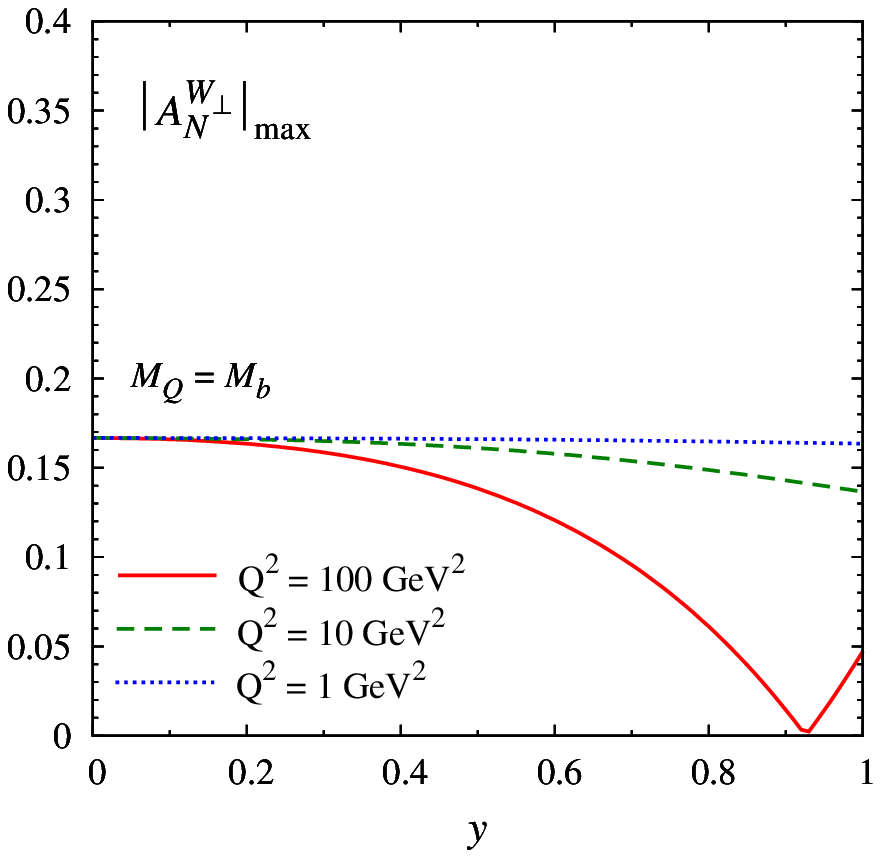}
 \caption{Same as in Fig.~1, but for the upper bounds $\vert A_N^W\vert_{\rm max}$ 
 of $A_N^{\sin(\phi^{\perp}_S+\phi^{\perp}_\sT)}$ and $A_N^{\sin(\phi^{\perp}_S-3\phi^{\perp}_\sT)}$.}
 \label{fig:ANW_y_perp}
\end{center}
\end{figure}

On the other hand, in order to isolate the $\phi_S$ dependent terms related to the $T$-odd gluon TMDs, we define the azimuthal moments
\begin{align}
A_N^{W(\phi_S,\phi_\sT)} & \equiv 2\, \frac {\int \d \phi_\sT \,\d\phi_\perp\, W(\phi_S,\phi_\sT)\,\left [\d\sigma(\phi_S,\,\phi_\sT,\,\phi_\perp) - \d\sigma(\phi_S + \pi,\,\phi_\sT,\,\phi_\perp)\right ]}{\int \d \phi_\sT \,\d\phi_\perp\, \left [\d\sigma(\phi_S,\,\phi_\sT,\,\phi_\perp) + \d\sigma(\phi_S + \pi,\,\phi_\sT,\,\phi_\perp)\right ]} ~.
\label{eq:mom}
\end{align}
After integration over $\phi_\perp$, only three independent modulations are left:
$\sin(\phi_S-\phi_\sT)$, $\sin(\phi_S+\phi_\sT)$,  $\sin(\phi_S-3\phi_\sT)$, each one related to a different
$T$-odd gluon TMD. These angular structures and the TMDs they probe are very similar to the case of 
quark asymmetries in SIDIS, $e\, p^\uparrow \to e' \, h\, X$, with $\phi_T$  replaced by $\phi_h$~\cite{Boer:1997nt}. The same holds for the $\cos 2 (\phi_\sT-\phi_\perp)$ asymmetry in Eq.~(\ref{eq:lpg}). More explicitly, we obtain:
\begin{align}
A_N^{\sin(\phi_S-\phi_\sT)} & =  \frac{\vert \bm q_\sT\vert}{M_p}\, \frac{f_{1T}^{\perp\,g}(x,\bm q_\sT^2)}{f_1^g(x,\bm q_\sT^2)}\,,
\label{eq:A1}\\
A_N^{\sin(\phi_S+\phi_\sT)}  &  =
\frac{\vert \bm q_\sT\vert}{M_p}\, \frac{h_{1}^{g}(x,\bm q_\sT^2)}{f_1^g(x,\bm q_\sT^2)}\,{\cal F}^{\sin(\phi_S+\phi_\sT)} (y, z,Q^2,M_\perp^2)  \,,
 \label{eq:A2}\\
A_N^{\sin(\phi_S-3\phi_\sT)}  &  = -  \frac{\vert \bm q_\sT\vert^3}{2 M_p^3}\, \frac{h_{1T}^{\perp \,g}(x,\bm q_\sT^2)}{f_1^g(x,\bm q_\sT^2)} \,   {\cal F}^{\sin(\phi_S-3\phi_\sT)} (y, z,Q^2,M_\perp^2)\,,   \label{eq:A3}
\end{align}
where ${ {\cal F}^{\sin(\phi_S+\phi_\sT)} =  {\cal F}^{\sin(\phi_S-3\phi_\sT)}= \cal F}^{\cos2\phi_\sT} \equiv {\cal F}$. 
Note that  $\langle \cos2\phi_\sT\rangle $, $A_N^{\sin(\phi_S+\phi_\sT)}$ and $A_N^{\sin(\phi_S-3\phi_\sT)}$ vanish in the limit $y\to 1$, which for $s/Q^2 \to \infty$ corresponds to the limit $x \to 0$, since  ${\cal F}\to 0$ when  $y\to 1$. Furthermore, a measurement of the ratio
\begin{equation}
\frac{A_N^{\sin(\phi_S-3\phi_\sT)}}{A_N^{\sin(\phi_S+\phi_\sT)}} =- \frac{\bm q_\sT^2}{2 M_p^2}\, \frac{h_{1T}^{\perp\,g}(x,\bm q_\sT^2)}{h_{1}^{g}(x,\bm q_\sT^2)}
\label{eq:ratioA}
\end{equation}
would probe directly the relative magnitude of $h_{1T}^{\perp\,g}$ and $h_1^g$ without dependences on any of the other kinematic variables in the process. From the positivity bounds in Eqs.~(\ref{eq:bound}), we find that the maximal value of the Sivers asymmetry in Eq.~(\ref{eq:A1}) is always one, while $\langle \cos 2\phi_\sT\rangle$ and the asymmetries in Eqs. (\ref{eq:A2}) and (\ref{eq:A3}) have the same upper bounds, that we denote by $A_N^W$. The maximal values of  $\vert A_N^W\vert $ are shown in Fig.~\ref{fig:ANW_y} as a function of $y$ in the kinematic region specified in the caption.
In this and subsequent numerical analyses we take $M_c= 1.3$ GeV and $M_b= 4.2$ GeV.

Alternatively,  the azimuthal angles can be defined w.r.t.\ $\phi_\perp$ instead of $\phi_\ell$, that can be integrated over. 
In analogy to Eqs.~(\ref{eq:A1})-(\ref{eq:A3}), we get
\begin{align}
A_N^{\sin(\phi_S^\perp-\phi_\sT^\perp)}  =  \frac{\vert \bm q_\sT\vert}{M_p}\, \frac{f_{1T}^{\perp\,g}}{f_1^g}\,,\quad
A_N^{\sin(\phi_S^\perp+\phi_\sT^\perp)}   =
\frac{\vert \bm q_\sT\vert}{M_p}\, \frac{h_{1}^{g}}{f_1^g}\,{\cal F}^{\perp}  \,,\quad
A_N^{\sin(\phi_S^\perp-3\phi_\sT^\perp)}   = -  \frac{\vert \bm q_\sT\vert^3}{2 M_p^3}\, \frac{h_{1T}^{\perp \,g}}{f_1^g} \,   {\cal F}^{\perp} \,,  
\end{align}
where ${\cal F^\perp}= {\cal F}^{\cos2(\phi_\sT-\phi_\perp)}$, see Eq.~(\ref{eq:lpg}). As a consequence, $\langle \cos2(\phi_\sT-\phi_\perp )\rangle $  has  the same upper bound of 
$A_N^{\sin(\phi_S^\perp+\phi_\sT^\perp)}$  and $A_N^{\sin(\phi_S^\perp+\phi_\sT^\perp)}$, named
$A_N^{W_\perp}$ and presented in Fig~\ref{fig:ANW_y_perp}. We note that the upper bounds of $A_N^{W_\perp}$ do not vanish when $y \to 1$, therefore these asymmetries are more suitable than $A_N^{W}$ for small-$x$ studies.

\begin{figure*}[t]
\begin{center}
\includegraphics[angle=0,width=0.48\textwidth]{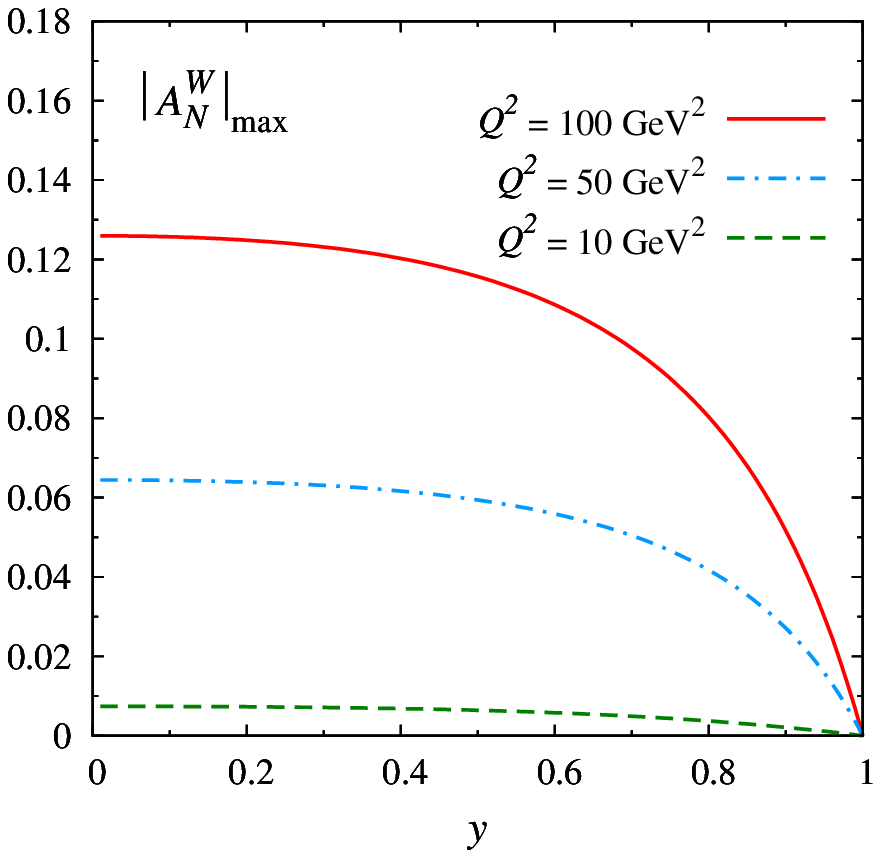}
 \includegraphics[angle=0,width=0.48\textwidth]{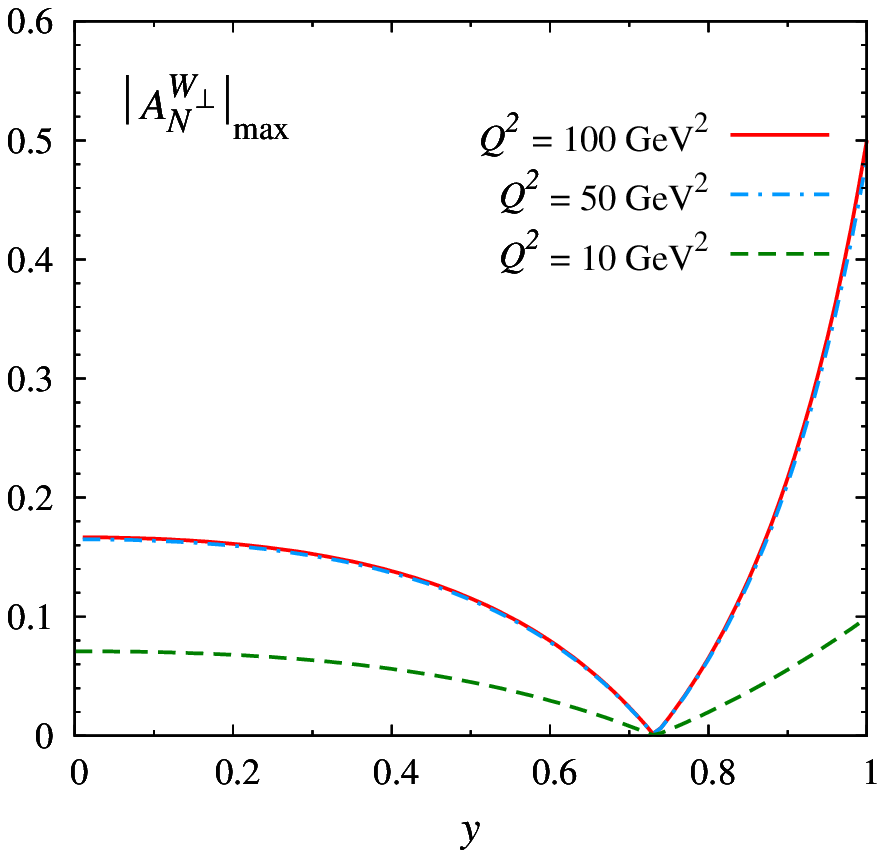}
 \caption{Upper bounds $\vert A_N^W\vert_{\rm max}$, with $W={\sin(\phi_S+\phi_\sT)},\, {\sin(\phi_S-3\phi_\sT)}$  (left panel), and  $\vert A_N^{W_\perp}\vert_{\rm max}$, with $W_\perp ={\sin(\phi^\perp_S+\phi^\perp_\sT)},\, {\sin(\phi^\perp_S-3\phi^\perp_\sT)}$ (right panel),  for the process $e p\to e^\prime {\rm{jet}}\, \rm{jet}\, X$. Estimates are given as a function of $y$, at different values of $K_\perp \ge 4$ GeV  (chosen in such a way that the asymmetries are maximal)  and $Q^2$, with $z=0.5$.}
\label{fig:ANW-y-dijet}
\end{center}
\end{figure*}

Turning now to the process $e(\ell) + p(P,S) \to e(\ell^{\prime}) + {\rm jet}(K_1) + {\rm jet}(K_2)+ X$, its cross section
at LO receives contributions from two partonic channels: $eq \to e^\prime q g$ and
$e g \to e^\prime q \bar q$. The angular structure of the cross section is the same as the one for heavy quark pair production given in Eq.~(\ref{eq:cs}). In a kinematic region where $x$ is small enough, the quark contribution can be neglected and the maximal values of the asymmetries for dijet production will be the same as the ones for heavy quark pair production in the limit of massless quarks.  These upper bounds for the $T$-odd gluon TMD asymmetries are shown in Fig.~\ref{fig:ANW-y-dijet}, where we have selected $K_\perp \ge 4$ GeV~\cite{Boer:2016fqd}.

\section{Process dependence of gluon TMDs}
The subprocesses $\gamma^*\, g \to Q \overline{Q}$ and $\gamma^*\, g \to q \bar{q}$, contributing respectively to heavy quark pair and dijet production, probe the gluon correlator in Eq.~(\ref{eq:PhiparT}) with two future pointing 
gauge links, which are referred
to as $+$ links as well.  The TMDs extracted from these processes can be related to the ones probed in specific $pp$ reactions. For instance,  in $p \, p\to \gamma \, \gamma \, X$~\cite{Qiu:2011ai} one can access gluon TMDs with
two past-pointing Wilson lines. Therefore, we predict the following relation for the gluon Sivers functions:
\begin{equation}
f_{1T}^{\perp\, g \, [e\, p^\uparrow \to e' \, Q \overline{Q}\, X]}(x,p_\sT^2) = - f_{1T}^{\perp\, g \, [p^\uparrow\,  p\to \gamma \, \gamma \, X]}
(x,p_\sT^2),
\end{equation}
and the same holds for the other two T-odd gluon TMDs, $h_1^g$ (or $h_{1T}$) and $h_{1T}^\perp$. This can be considered as the gluon analogue of the sign change relation between the quark Sivers function probed in SIDIS and Drell-Yan. In contrast, since $T$-even gluon TMDs with two $-$ links are equal to those with two $+$ links, it should be
\begin{equation}
h_{1}^{\perp\, g \, [e\, p \to e' \, Q \overline{Q}\, X]}(x,p_\sT^2) = h_{1}^{\perp\, g \, [p\,  p\to \gamma\, \gamma \, X]}
(x,p_\sT^2).
\end{equation}
Similar conclusions can be drawn if, instead of $\gamma\,\gamma$, one considers any other color-singlet final state in
$gg$-dominated kinematics, such as  a $J/\psi-\gamma$ pair~\cite{Dunnen:2014eta}, a (pseudo)scalar quarkonium~\cite{Boer:2012bt}, or even a Higgs boson~\cite{Boer:2011kf}.  This provides additional motivation to study gluon TMDs at RHIC or AFTER@LHC and compare them to EIC studies in the future.

On the other hand, by looking at the process $p\, p\to \gamma \, \text{jet}\, X$ in the kinematic regime dominated by gluons inside one of the protons,  the subprocess $q \, g \to \gamma \, q$ can be selected, which probes gluon TMDs with future and past
pointing Wilson lines, {\it i.e.}\ a  $+$ and $-$ link. Hence the two processes $e\, p \to e' \, Q \overline{Q}\, X$ and $p\, p\to \gamma \, \text{jet}\, X$  probe distinct and entirely independent gluon TMDs, which cannot be related to each other. We conclude that TMD observables  accessible at the LHC and at EIC can be either related or complementary depending on the process.

\section{Gluon TMDs and azimuthal asymmetries in the small-$x$ region}
\begin{figure}[t]
\centering
\includegraphics[angle=0,width=0.45\textwidth]{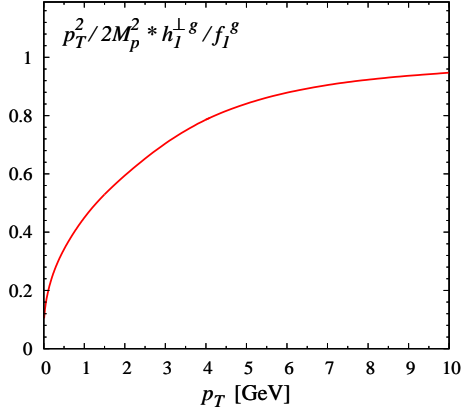}
\caption{Ratio of the WW-type gluon TMDs $h_1^{\perp\,g}$ and $f_1^g$ in the MV model as a function of $p_\sT$.}
\label{fig:bound-mv}
\end{figure}
\begin{figure}[t]
\begin{center}
 \includegraphics[angle=0,width=0.48\textwidth]{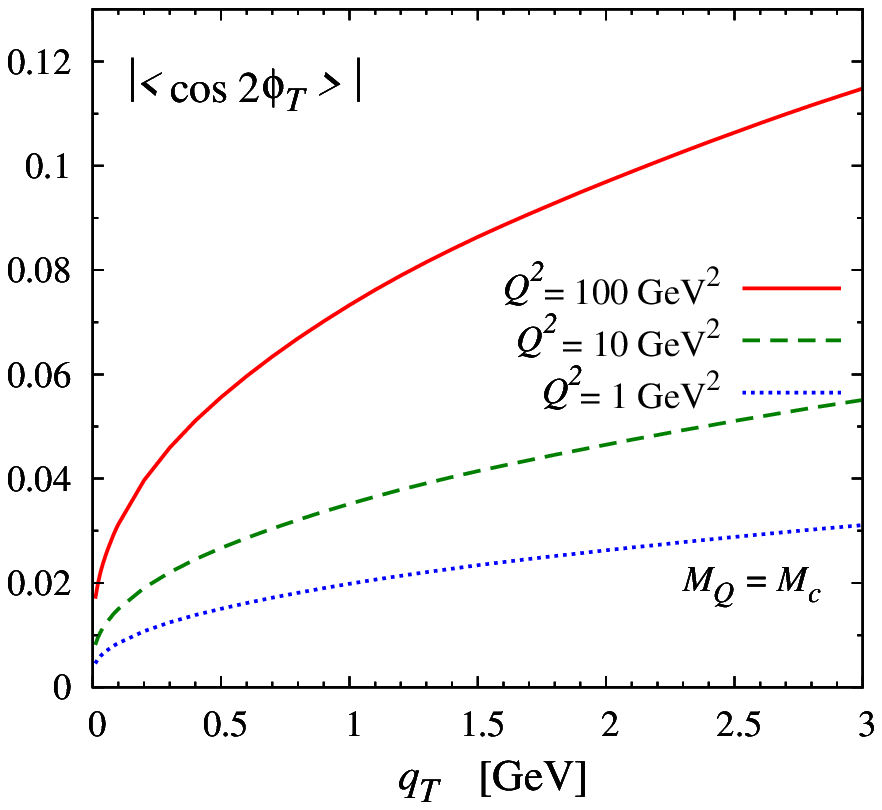}
 \includegraphics[angle=0,width=0.48\textwidth]{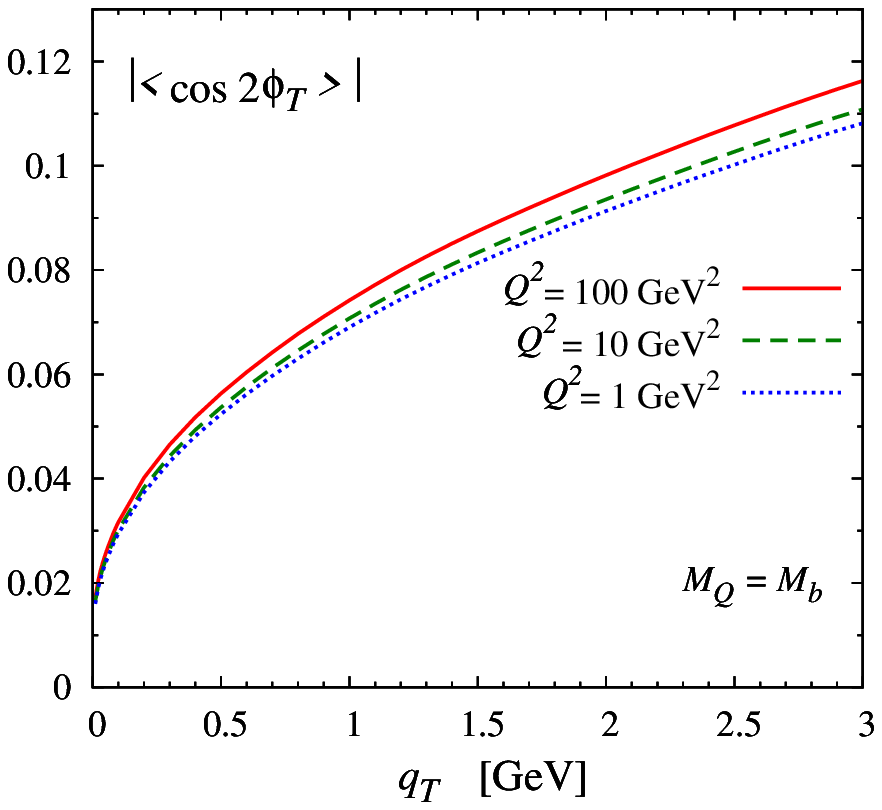}
 \caption{$\vert \langle\cos 2 \phi_\sT \rangle \vert $ asymmetry in the MV model, as a function of $q_\sT \equiv \vert \boldsymbol q_\sT\vert$, calculated at $ \vert \boldsymbol K_\perp \vert = 6 $ GeV,  $z=0.5$,  $y=0.1$ and at different values of  $Q^2$,  for  $e p\to e^\prime c \,\bar{c} \, X$  (left panel) and   $e p\to e^\prime b \,\bar{b} \, X$  (right panel).}
\label{figRp}
\end{center}
\end{figure}
\begin{figure}[t]
\begin{center}
 \includegraphics[angle=0,width=0.48\textwidth]{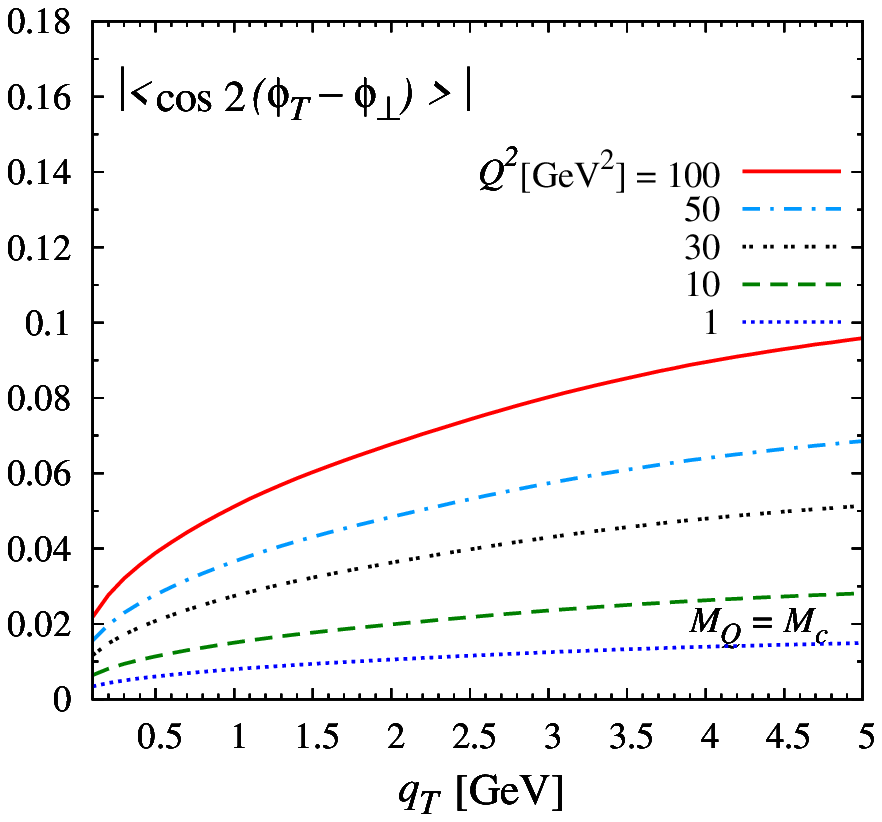}
 \includegraphics[angle=0,width=0.48\textwidth]{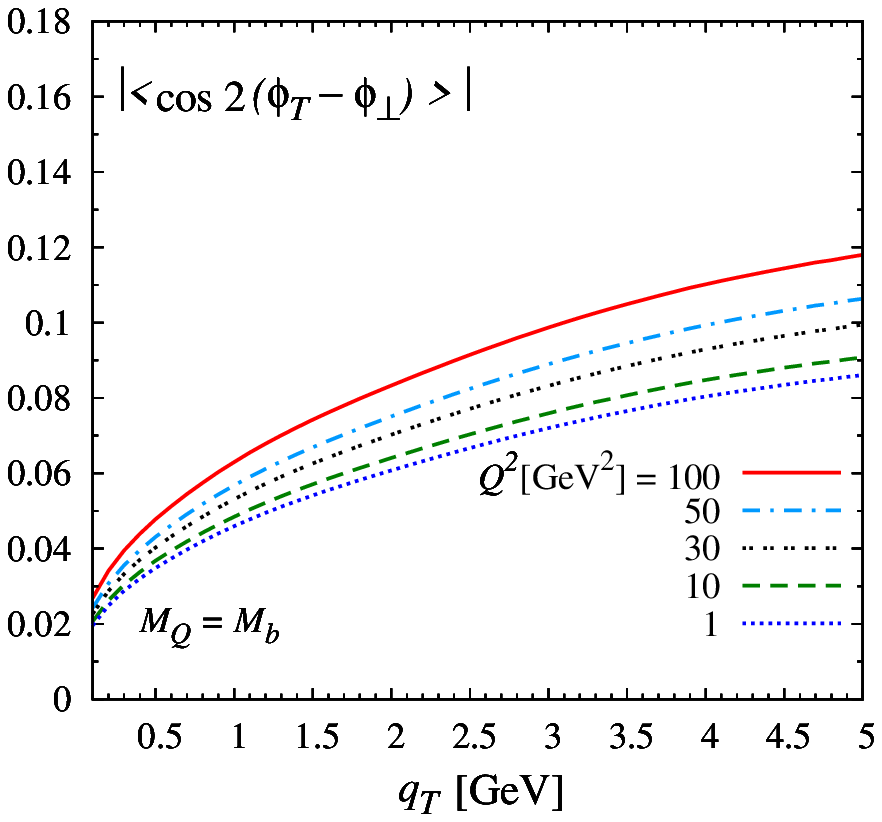}
\caption{ $\vert \langle\cos 2 (\phi_\sT-\phi_\perp) \rangle \vert $ asymmetry in the MV model, as a function of $q_\sT \equiv \vert \boldsymbol q_\sT\vert$, calculated at $ \vert \boldsymbol K_\perp \vert = 10 $ GeV,  $z=0.5$, $y=0.3$,  and different values of  $Q^2$, for $e p\to e^\prime c \,\bar{c} \, X$ (left panel) and  $e p\to e^\prime b \,\bar{b} \,X$ (right panel).}
\label{figR}
\end{center}
\end{figure}

The gluon TMDs that can be accessed in heavy quark and dijet production in DIS, characterized by two + gauge links, correspond to the so-called Weisz\"acker-Williams (WW) distributions at small $x$~\cite{Dominguez:2011wm,Dominguez:2010xd}, while those with a $+$ and a $-$ link correspond to dipole distributions. Both the WW and the dipole linearly polarized gluon distributions can be calculated using saturation models, since they have the same $\ln 1/x$ enhancement as the unpolarized TMDs in the small-$x$ region. For example, the ratio between the WW linearly polarized and unpolarized gluon TMDs in a McLerran-Venugopalan (MV) model~\cite{McLerran:1993ni}  is given by~\cite{Metz:2011wb,Dominguez:2011br,Boer:2016fqd}
\begin{equation}
\frac{\bm p_\sT^2 }{2M_p^2}\,\frac{h_1^{\perp\,g}(x, \bm p_\sT^2)}{f_1^g(x, \bm p_\sT^2)} = \frac{
\int_0^\infty \frac{\d r_\perp}{r_\perp}\,  J_2 \left (| \bm p_\sT| r_\perp \right ) {{\rm ln}^{-1}
\left [ \frac{1}{r_\perp^2 \Lambda^2_{QCD}}+e\right ]}\, \left \{ 1-e^{ - \frac{r_\perp^2
Q_{s0}^2}{4}{\rm ln} \left [ \frac{1}{r^2_\perp \Lambda^2_{QCD} }+e \right ]} \right
\}}{\int_0^\infty \frac{\d r_\perp}{r_\perp}\,  J_0 \left (| \bm p_\sT| r_\perp \right ) \, \left
\{ 1-e^{ - \frac{r_\perp^2 Q_{s0}^2}{4}{\rm ln} \left [ \frac{1}{r_\perp^2 \Lambda^2_{QCD} }+e
\right ]} \right \}}\,,
\label{eq:ratioMV}
\end{equation}
where $Q_{s0}$ is the saturation scale, which has to be extracted from fits to experimental data. This ratio is shown in Fig.~\ref{figRp} for the $Q_{s0}^2= 0.35$ GeV$^2$ at $x=0.01$ and $\Lambda_{QCD} = 0.2$ GeV.  It turns out that this ratio  is not saturated for $p_\sT$ values below 10 GeV, although it is still sizable except for very small $p_\sT$ values. This is in constrast  with the dipole case, in which such ratio is always saturated~\cite{Metz:2011wb}. Using the MV model expression in Eq.~(\ref{eq:ratioMV}), we show in Figs.~\ref{figRp} and \ref{figR} our estimates for the absolute values of $\langle \cos2\phi_T\rangle $ and 
$\langle \cos2(\phi_T-\phi_\perp) \rangle $, respectively,  in two different kinematic regions. Both asymmetries are sizable and decrease for values of $y$ and $\vert \bm K_\perp \vert$ larger than the ones chosen. 
Note that  the center of mass energy has to be sufficiently large for these curves to be in the small-$x$ range. For an EIC, the planned $\sqrt{s}$ varies from about 20 to 150 GeV \cite{Boer:2011fh}, hence $x_B = Q^2/ys<0.01$ for $y\geq 0.1$ requires $Q^2 < (10^{-3}-10^{-2}) s$. For $\sqrt{s}=60$ GeV  the $Q^2=100$ GeV$^2$ curve should not be considered. The large $Q^2$ curves are included for a high energy EIC or for an analysis of HERA data, for which  $\sqrt{s}=320$ GeV.

Turning to the small-$x$ behavior of the WW-type $T$-odd gluon TMDs, this can be perturbatively computed at large $p_\sT$ following the standard collinear twist-3 approach~\cite{Boer:2015pni}. Such studies suggest that the WW $T$-odd distributions are  suppressed by a factor of $x$ as compared to the corresponding dipole ones. The complete expressions for their sub-asymptotic behavior, which  cannot be estimated in the MV model, are given in Ref.~\cite{Boer:2016fqd}, where three
different models for the WW gluon TMDs have also been proposed. The ratio of asymmetries in Eq.~(\ref{eq:ratioA}) could be used to test such models, because each one of them leads to a different prediction for the ratio.   

\section{Conclusions}

Azimuthal asymmetries in heavy quark pair and dijet production in DIS processes could provide direct access to  gluon TMDs.  These asymmetries are very similar to the ones measured in semi-inclusive DIS in order to extract quark TMDs.
We have identified the kinematic regions in which they are  maximal and found that, in such domains, the corresponding upper bounds are sizable.  Their measurements at an EIC would allow to test our prediction of the sign change of the $T$-odd gluon TMDs by comparison  with the corresponding observables at RHIC and AFTER$@$LHC. 

In the small-$x$ region, the TMDs under investigation correspond to the WW-type gluon distributions. Although the WW linearly polarized gluon distribution, unlike the dipole one, does not saturate the positivity bound, its effects in heavy quark pair and dijet production at an EIC should still be measurable, as it turns out from our analysis in the framework of the MV model. Saturation models cannot be used to estimate the $T$-odd WW gluon TMDs because their asymptotic behavior at small-$x$ vanishes. Nevertheless,  we propose to look at the  ratio of the $\sin(\phi_S+\phi_\sT)$ and $\sin(\phi_S-3\phi_\sT)$ modulations, for which other model expectations are provided~\cite{Boer:2016fqd}, to probe two of the $T$-odd TMDs.  It will be interesting to check
experimentally  the theoretical prediction of  different asymptotic behavior  at small $x$ of  the WW and dipole type $T$-odd gluon TMDs. The latter ones could be probed in other processes, such as virtual photon plus jet production in polarized proton-proton collisions at RHIC. To conclude, the studies of azimuthal asymmetries in heavy quark pair and dijet  production could form a prominent part of both the spin and the small-$x$ physics programs at a future EIC.

\acknowledgments
We acknowledge the financial support of the European Research Council (ERC) under the FP7 ``Ideas'' programme (grant agreement No.~320389, QWORK) and the European Union's Horizon 2020 research and innovation programme (grant agreement No. 647981, 3DSPIN).


\begin{thebibliography}{99}

\bibitem{Mulders:2000sh}
  P.~J.~Mulders and J.~Rodrigues,
  \emph{Transverse momentum dependence in gluon distribution and fragmentation functions},
  Phys.\ Rev.\  D {\bf 63}   (2001) 094021
 [ {\tt arXiv:hep-ph/0009343}].

\bibitem{Sivers:1989cc}
  D.~W.~Sivers,
  \emph{Single Spin Production Asymmetries from the Hard Scattering of Point-Like Constituents},
  \emph{Phys.\ Rev.\ D} {\bf 41} (1990) 83.

\bibitem{Boer:2015ika}
  D.~Boer, C.~Lorc\'e, C.~Pisano and J.~Zhou,
  \emph{The gluon Sivers distribution: status and future prospects},
  \emph{Adv.\ High Energy Phys.}\  {\bf 2015} (2015) 371396
  [{\tt arXiv:1504.04332 [hep-ph]}].

\bibitem{Boer:2016fqd}
  D.~Boer, P.~J.~Mulders, C.~Pisano and J.~Zhou,
  \emph{Asymmetries in Heavy Quark Pair and Dijet Production at an EIC},
  \emph{JHEP} {\bf 1608} (2016) 001
  [{\tt arXiv:1605.07934 [hep-ph]}].

\bibitem{Boer:2010zf}
  D.~Boer, S.~J.~Brodsky, P.~J.~Mulders and C.~Pisano,
  \emph{Direct Probes of Linearly Polarized Gluons inside Unpolarized Hadrons,
  Phys.\ Rev.\ Lett.\  {\bf 106}, 132001 (2011)}
  [{\tt arXiv:1011.4225 [hep-ph]}].

\bibitem{Pisano:2013cya}
  C.~Pisano, D.~Boer, S.~J.~Brodsky, M.~G.~A.~Buffing and P.~J.~Mulders,
  \emph{Linear polarization of gluons and photons in unpolarized collider experiments},
  \emph{JHEP} {\bf 1310}, 024 (2013)
  [{\tt arXiv:1307.3417 [hep-ph]}].

\bibitem{Boer:2011fh}
  D.~Boer {\it et al.},
  \emph{Gluons and the quark sea at high energies: Distributions, polarization, tomography},
  {\tt arXiv:1108.1713 [nucl-th]}.
  
\bibitem{Brodsky:2012vg}
  S.~J.~Brodsky, F.~Fleuret, C.~Hadjidakis and J.~P.~Lansberg,
  \emph{Physics Opportunities of a Fixed-Target Experiment using the LHC Beams},
  \emph{Phys.\ Rept.}\  {\bf 522} (2013) 239
  [{\tt arXiv:1202.6585 [hep-ph]}].

\bibitem{Meissner:2007rx}
  S.~Meissner, A.~Metz and K.~Goeke,
  \emph{Relations between generalized and transverse momentum dependent parton distributions},
  Phys.\ Rev.\ D {\bf 76} (2007) 034002 
  [{\tt hep-ph/0703176 [HEP-PH]}].

\bibitem{Boer:1997nt} 
  D.~Boer and P.~J.~Mulders,
  \emph{Time reversal odd distribution functions in leptoproduction},
  \emph{Phys.\ Rev.\ D} {\bf 57} (1998) 5780.
  [{\tt hep-ph/9711485}].

\bibitem{Qiu:2011ai}
  J.~W.~Qiu, M.~Schlegel and W.~Vogelsang,
  \emph{Probing Gluonic Spin-Orbit Correlations in Photon Pair Production},
  \emph{Phys.\ Rev.\ Lett.}\  {\bf 107} (2011) 062001
  [{\tt arXiv:1103.3861 [hep-ph]}].

\bibitem{Dunnen:2014eta}
  W.~J.~den Dunnen, J.~P.~Lansberg, C.~Pisano and M.~Schlegel,
  \emph{Accessing the Transverse Dynamics and Polarization of Gluons inside the Proton at the LHC},
  \emph{Phys.\ Rev.\ Lett.}\  {\bf 112} (2014)  212001
  [{\tt arXiv:1401.7611 [hep-ph]}].

\bibitem{Boer:2012bt}
  D.~Boer and C.~Pisano,
  \emph{Polarized gluon studies with charmonium and bottomonium at LHCb and AFTER},
  \emph{Phys.\ Rev.\ D} {\bf 86} (2012) 094007
  [{\tt arXiv:1208.3642 [hep-ph]}].
  
\bibitem{Boer:2011kf}
  D.~Boer, W.~J.~den Dunnen, C.~Pisano, M.~Schlegel and W.~Vogelsang,
  \emph{Linearly Polarized Gluons and the Higgs Transverse Momentum Distribution},
  \emph{Phys.\ Rev.\ Lett.\ } {\bf 108} (2012) 032002 
  [{\tt arXiv:1109.1444 [hep-ph]}].

\bibitem{Dominguez:2011wm}
  F.~Dominguez, C.~Marquet, B.~W.~Xiao and F.~Yuan,
  \emph{Universality of Unintegrated Gluon Distributions at small x},
  \emph{Phys.\ Rev.\ D} {\bf 83} (2011) 105005 
  [{\tt arXiv:1101.0715 [hep-ph]}].

\bibitem{Dominguez:2010xd}
  F.~Dominguez, B.~W.~Xiao and F.~Yuan,
  \emph{$k_t$-factorization for Hard Processes in Nuclei},
  \emph{Phys.\ Rev.\ Lett.}\  {\bf 106} (2011) 022301
   [{\tt arXiv:1009.2141 [hep-ph]}].

\bibitem{McLerran:1993ni}
  L.~D.~McLerran and R.~Venugopalan,
  \emph{Computing quark and gluon distribution functions for very large nuclei},
  \emph{Phys.\ Rev.\ D} {\bf 49} (1994) 2233 
  [{\tt hep-ph/9309289}].

\bibitem{Metz:2011wb}
  A.~Metz and J.~Zhou,
  \emph{Distribution of linearly polarized gluons inside a large nucleus},
  \emph{Phys.\ Rev.\ D} {\bf 84} (2011) 051503
  [{\tt arXiv:1105.1991 [hep-ph]}].

\bibitem{Dominguez:2011br}
  F.~Dominguez, J.~W.~Qiu, B.~W.~Xiao and F.~Yuan,
  \emph{On the linearly polarized gluon distributions in the color dipole model},
  \emph{Phys.\ Rev.\ D}  {\bf 85} (2012)  045003 
  [{\tt arXiv:1109.6293 [hep-ph]}].
 

\bibitem{Boer:2015pni}
  D.~Boer, M.~G.~Echevarria, P.~Mulders and J.~Zhou,
  \emph{Single spin asymmetries from a single Wilson loop},
  \emph{Phys.\ Rev.\ Lett.}\  {\bf 116} (2016) 122001
  [{\tt arXiv:1511.03485 [hep-ph]}].


\end{thebibliography}
\end{document}